\documentclass[sigconf]{acmart}
\usepackage{balance}
\usepackage{hyperref}
\usepackage{graphicx}
\usepackage{booktabs} % For formal tables
\usepackage{amsmath}
\usepackage{amsfonts}
\usepackage{amsthm}
\usepackage{algorithm}
\usepackage{algpseudocode}
\usepackage{subfigure}
\usepackage{amsfonts}
\usepackage{mdwlist}
\usepackage{gensymb}
\usepackage[hang,flushmargin]{footmisc}
\usepackage{xcolor}
\usepackage{appendix}
\usepackage{makecell}

\newcommand{\mwtech}{multi-streaming}
\newcommand{\mwpfull}{Multi-Stream Party}
\newcommand{\mwp}{MSP}
\newcommand{\tensorfull}{\underline{S}ocio-t\underline{e}mporal Do\underline{n}ation-Re\underline{s}ponse Tens\underline{or} Co-factorization}
\newcommand{\tensor}{SENSOR}
\newcommand{\frameworkfull}{\underline{M}ulti-stream P\underline{a}rty \underline{R}ecommender \underline{S}ystem}
\newcommand{\framework}{MARS}
\newcommand{\rankf}{CARS-F}
\newcommand{\ranku}{CARS-U}
\newcommand{\rankc}{CARS-C}
\newcommand{\rankn}{CARS-N}
\newcommand{\ripple}{STAR}
\newcommand{\smooth}{RIOT}
\newcommand{\rankfull}{\underline{C}hannel Influence-\underline{A}ware \mwp\ \underline{R}anking \underline{S}ystem}
\newcommand{\rank}{CARS}
\newcommand{\probfull}{\underline{D}onation \underline{a}nd \underline{M}SP \underline{Rec}ommendation}
\newcommand{\prob}{DAMRec}
\newcommand{\dtor}{D2R}
\newcommand{\dtorn}{D2R-N}
\newcommand{\tensorser}{SENSOR-nSER}
\newcommand{\tensors}{SENSOR-nSTAR}
\newcommand{\tensorr}{SENSOR-nRIOT}
\newcommand{\tensorn}{SENSOR-Na\"{i}ve}
\newcommand{\ser}{SER}

\copyrightyear{2020}
\acmYear{2020}
\setcopyright{acmcopyright}\acmConference[CIKM '20]{Proceedings of the 29th ACM International Conference on Information and Knowledge Management}{October 19--23, 2020}{Virtual Event, Ireland}
\acmBooktitle{Proceedings of the 29th ACM International Conference on Information and Knowledge Management (CIKM '20), October 19--23, 2020, Virtual Event, Ireland}
\acmPrice{15.00}
\acmDOI{10.1145/3340531.3411925}
\acmISBN{978-1-4503-6859-9/20/10}

\begin{document}
\fancyhead{}

\title[Donation and Multi-stream Recommendation]{Live Multi-Streaming and Donation Recommendations via Coupled Donation-Response Tensor Factorization}

\author{Hsu-Chao Lai}
\affiliation{
  \institution{Dept. of Computer Science, National Chiao Tung Univ.}
}
\affiliation{
  \institution{Inst. of Information Science, Academia Sinica, Taiwan}
  %\country{Taiwan}
}
\email{hsuchao.cs05g@nctu.edu.tw}
\author{Jui-Yi Tsai}
\affiliation{
  \institution{Dept. of Electrical and Computer Engineering, National Chiao Tung Univ.}
}
\affiliation{
  \institution{Inst. of Information Science, Academia Sinica, Taiwan}
  %\country{Taiwan}
}
\email{vincenttsai@iis.sinica.edu.tw}
\author{Hong-Han Shuai}
\affiliation{%
  \institution{Dept. of Electrical and Computer Engineering, National Chiao Tung Univ., Taiwan}
  %\country{Taiwan}
}
\email{hhshuai@nctu.edu.tw}

\author{Jiun-Long Huang}
\affiliation{
  \institution{Dept. of Computer Science, National Chiao Tung Univ., Taiwan}
  %\country{Taiwan}
}
\email{jlhuang@cs.nctu.edu.tw}

\author{Wang-Chien Lee}
\affiliation{
  \institution{Dept. of Computer Science and Engineering, The Pennsylvania State Univ., USA}
  %\country{USA}
}
\email{wlee@cse.psu.edu}
\author{De-Nian Yang}
\affiliation{
  \department{Inst. of Information Science}
  \department{Research Center for Information Technology Innovation}
  \institution{Academia Sinica, Taiwan}
  %\country{Taiwan}
}
\email{dnyang@iis.sinica.edu.tw}

\begin{abstract}
In contrast to traditional online videos, live multi-streaming supports real-time social interactions between multiple streamers and viewers, such as donations. However, donation and multi-streaming channel recommendations are challenging due to complicated streamer and viewer relations, asymmetric communications, and the tradeoff between personal interests and group interactions. In this paper, we introduce \textit{\underline{M}ulti-\underline{S}tream \underline{P}arty (MSP)} and formulate a new multi-streaming recommendation problem, called \textit{\underline{D}onation \underline{a}nd \underline{M}SP \underline{Rec}ommendation (DAMRec)}. We propose \textit{\underline{M}ulti-stream P\underline{a}rty \underline{R}ecommender \underline{S}ystem (MARS)} to extract latent features via socio-temporal coupled donation-response tensor factorization for donation and MSP recommendations. Experimental results on Twitch and Douyu manifest that MARS significantly outperforms existing recommenders by at least 38.8\% in terms of hit ratio and mean average precision.
\end{abstract}

\maketitle

\section{Introduction}
Live streaming platforms, such as Twitch, Facebook Gaming, YouTube Live, and  Microsoft Mixer, which broadcast music, political, and gaming content to viewers, have recently grown into one of the most popular social services. For Twitch, the number of daily active viewers grows up to 10 million \cite{donate18CHI}, the minutes watched shoot up to 434 billion,\footnote{\url{https://bit.ly/2mBYCcq}} and the revenue from advertisements and donations hits 545 billion in 2018.\footnote{\url{https://bit.ly/2Ik9uVd}}

The great success of live streaming may be credited to several unique functions. First, viewers may interact with other viewers and streamers in real-time. Friends interact 10 times more on Facebook Live than in regular videos.\footnote{\url{https://bit.ly/1V9oxkl}} More than 21\% of friends in bilibili simultaneously watch the same videos and chat with more than 26.5 lines on a 3-minute video. Therefore, live streaming platforms have been recognized as the "third places" to enjoy social interactions and form communities \cite{donate18china,donate18CHI}. Indeed, social interactions are important to bring friends together to enjoy real-time live streaming. While \textit{group watching} is growing with great momentum, existing live streaming platforms recommend channels to each viewer by conventional personalized recommendation \cite{ChenZ0NLC17,LiuCLH19}, which fails to foster social interactions to improve viewer engagement. To stimulate discussions, a simple way is to adopt group recommenders \cite{DC18,vldb20vr} to find a unified set of channels for a group of friends. Nevertheless, the recommended channels are identical for each viewer, which fails to address diverse personal interests.

\begin{figure}[tp]
	\centering
	\subfigure[][Multi-stream on Twitch.] {\ \centering \includegraphics[width = 0.47 \columnwidth] {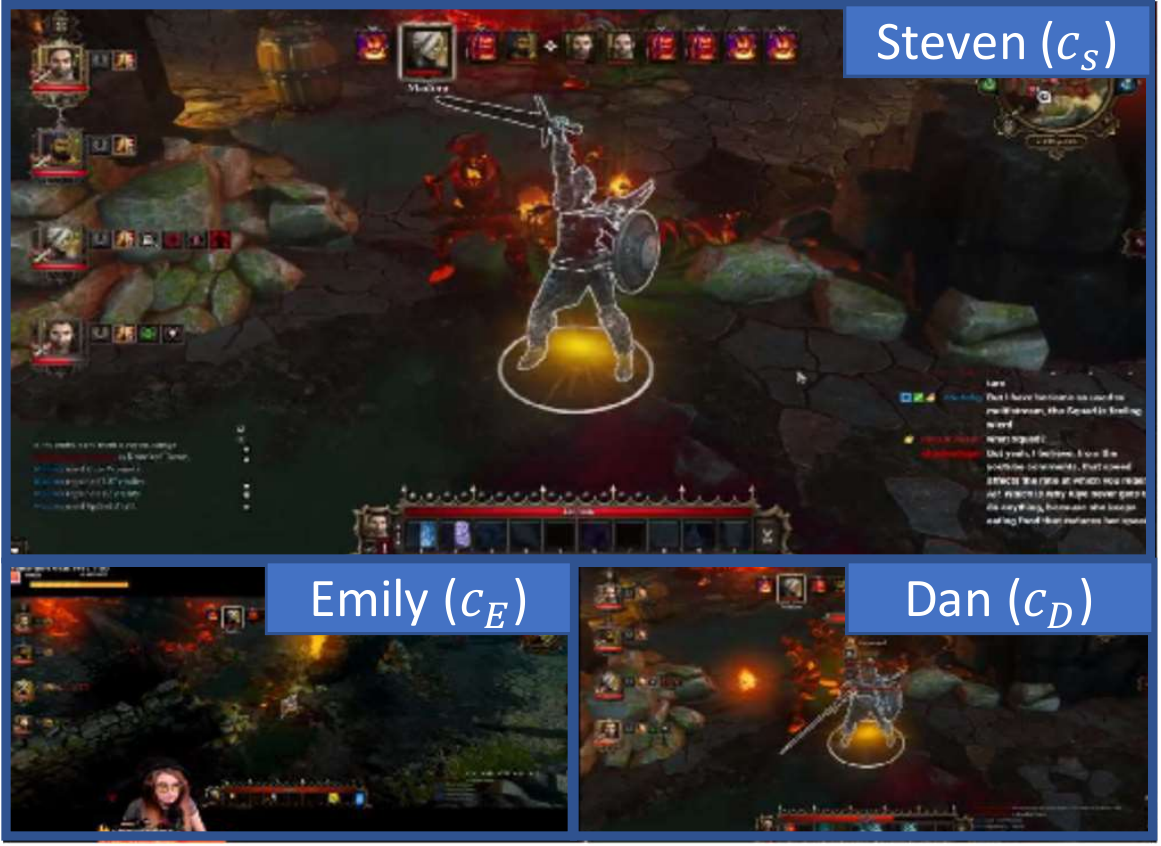}
		\label{fig:squad} } 
	\subfigure[][An \mwp\ example.] {\	\centering \includegraphics[width = 0.47 \columnwidth] {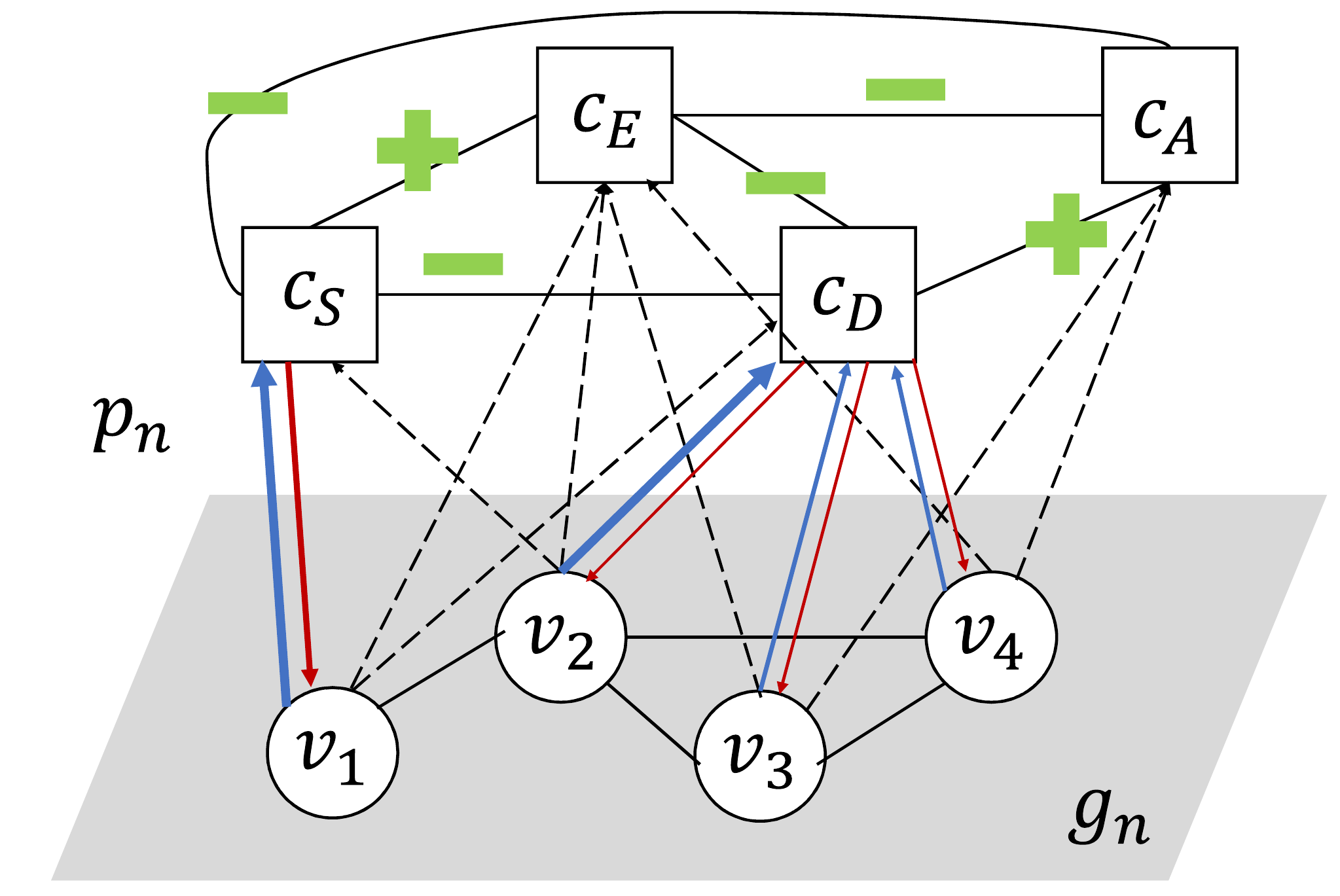}
		\label{fig:mwp1} }
	\caption{An illustrative example of \mwp.}
	\label{fig:MWP_example}
\end{figure}

Second, current live streaming platforms support viewers to watch multiple channels, i.e., \textit{multi-streams}, simultaneously. NBC and FOX offer at least four viewing angles in sports games (e.g., NFL, MLB, and UFC), attracting 25M and 48M viewers, respectively. The analyses on Twitch \cite{twitch_chat_data} and 2017 Taipei Summer Universiade on YouTube Live \cite{CYS18} manifest that 27\% and 30\% of viewers simultaneously watch multiple channels, respectively. A Twitch user study reveals that more than 90\% of users have enjoyed multi-streaming \cite{CYS18}. Figure \ref{fig:squad} shows the user interface of \textit{Squad Team} (a multi-streaming program on Twitch) with Steven and Emily as teammates and Dan as their opponents in a four-player game. More than 3K viewers were enjoying watching the collaboration and competition between streamers, indicating that streamer relations (e.g., teammates or opponents) and interactions are crucial in multi-streams. With multi-streaming, viewers can enjoy their favorite streamers and channels, while engaging in different communities. Currently, streamers on Twitch or Mixer have to manually select their own partners to broadcast multi-streams together instead of generating multi-streams via recommender systems to a group of viewers based on their preferences. 

Third, in contrast to conventional online video streaming \cite{ChenZ0NLC17,LiuCLH19}, live streaming supports \textit{donations}. Viewers donate, send gifts, and leave messages to the streamers while enjoying social interactions with others in the channels. Streamers usually respond immediately or host online events to reward donors. As a result, the viewers are more engaged in the channels and thereby more willing to donate (which profits the streamers and the platforms). Moreover, social interactions between viewers and friends (e.g., discussions on the game strategy for a common player) while watching a common channel also stimulate donations to attract and compete for the streamers' attention \cite{donate18CHI,donate18china}. Viewers sometimes compete for the title of top fans of a channel to impress the streamers and show their loyalty. While live streaming has provided single-channel recommendations to users \cite{ChenZ0NLC17,LiuCLH19}, it does not support donation recommendations, which could be a promising service. 

In this paper, we envisage a scenario, termed as \textit{\mwpfull\ (\mwp)}, where a group of friends watches multiple live streaming channels together with a satisfactory experience. In an \mwp, the recommended channels may vary for each viewer based on her personal interests. Figure \ref{fig:mwp1} illustrates an example of \mwp\ with the viewers (circles) and their social relationships (solid lines) presented at the bottom. Each viewer watches (blue and dashed arrows) three channels (squares) at the top. Streamer relations (solid lines) with green signs indicate their polarities (i.e., "+" for positive and "-" for negative relationships). Let channels $c_S$, $c_E$, and $c_D$ represent Steven, Emily, and Dan in Figure \ref{fig:squad}, and $c_A$ be the channel of Dan's teammate. The negative solid line indicates that $c_S$ and $c_D$ have a negative relationship (e.g., opponents). Viewers $v_1$ and $v_2$ watch a common channel $c_S$ to enjoy social interactions. However, viewers $v_3$ and $v_4$, not interested in $c_S$, watch $c_E$, $c_D$, and $c_A$ instead. As such, viewers are no longer tied to commonly selected channels. Moreover, $v_1$ donates to $c_S$ (Steven) for a nice play, and Steven replies with warm gratitude (thick red arrows). In contrast, $v_2$, $v_3$, and $v_4$ simultaneously donate to $c_D$ to encourage Dan when he is outplayed by Steven. However, since it is hard for Dan to properly reply to all donations during the game, the responses from Dan are brief (thin red arrows).

To meet the need of configuring a good \mwp, we formulate the problem of \textit{\probfull\ (\prob)}. An \mwp, representing a candidate plan of grouping viewers with channels, consists of $k$ channels for each viewer in a group, where the set of $k$ channels for each viewer is not necessarily identical. Given the viewer donations and streamer responses in the past, the first goal of \prob\ is to recommend a streamer to a viewer for donations to maximize the expected \textit{reciprocal response}. In Twitch, Douyu, and Mixer, streamers offer various responses, e.g., verbal sentences or activity invitations, to express their gratitude. Viewers are usually satisfied with high-quality reciprocal responses due to the senses of being valued. The second goal of \prob\ is to recommend an \mwp\ achieving high \textit{\mwp\ personal satisfaction}. 

The above two goals are entangled and thus need to be addressed together. If a viewer is not satisfied with the recommended \mwp, she is unlikely to enjoy those channels and donate to the streamers. However, if we recommend \mwp s first, the streamers of \mwp\ may not provide high-quality responses for donations when there are numerous viewers. On the other hand, prioritizing donations may favor some unpopular channels (where streamers offer more responses), which cannot fit the viewers' interests. 

Indeed, new challenges arise for \prob\ (detailed in Section \ref{sec:challenge}). Challenge 1 (C1): The aforementioned \textit{seesaw in optimizing \mwp\ and donation recommendations} needs to be solved. Challenge 2 (C2): Recommending donations for maximum reciprocal responses is challenging due to \textit{complicated streamer relations and viewer relations}, including streamer signed social relations and the socio-temporal ripple effect on donations, and Challenge 3 (C3) \textit{Asymmetric viewer and streamer communication behaviors}. Challenge 4 (C4) It is hard to quantify the personal satisfaction for recommending \mwp\ due to the \textit{tradeoff between personal interests and group interactions}.

\sloppy
In this paper, we propose a two-phase \textit{\frameworkfull\ (\framework)}. In phase 1, a new couple tensor factorization model \textit{\tensorfull\ (\tensor)} is proposed to jointly extract the latent representations (i.e., embeddings) of viewers and channels in order to capture socio-temporal behaviors of donations and responses. We introduce the \textit{\underline{D}onation \underline{to} \underline{R}esponse estimation (\dtor)} to estimate the reciprocal response of a given donation for donation recommendations. In phase 2, a new ranking method \textit{\rankfull\ (\rank)} is designed to rank \mwp s based on the viewer and channel embeddings derived from \tensor\ and the channel influence, which factors in the interplay among personal interests, group interactions, and streamer relations, for different candidate MSPs. The contributions are summarized as follows.

\begin{itemize}
    \item \mwp\ is introduced for live streaming viewers to enjoy social interactions while watching different preferred channels. We formulate \prob\ and make the first attempt to recommend donations and \mwp s.
    \item We propose a novel machine learning framework \framework\ for \prob. A new coupled tensor factorization model \tensor\ is introduced to extract the embeddings of viewers and channels. A set of regularizers are proposed to respectively account for streamer relations, the socio-temporal ripple effect of donations, and the asymmetric communications of streamers and viewers to accurately estimate the reciprocal responses to recommend donations.
    \item We design \rank\ to accurately rank candidate \mwp s for viewers according to channel influence, personal interests, viewer interactions, and streamer relations.
    \item We collect a new Twitch dataset. Large-scale experiments are conducted on four real datasets. Experimental results manifest that \dtor\ significantly outperforms feedback prediction models by at least 41.9\% regarding the root-mean-square error, and \rank\ outperforms personalized and group recommenders by at least 38.8\%  and 40.4\% regarding hit ratio and mean average precision, respectively.
\end{itemize}

\section{Related Work}
\textbf{Live Streaming Research.} 
Existing research focuses on enhancing watching experiences by optimizing transmitting latency and bitrate \cite{wc19acmmm}, while some research studies the splitting strategy of donations for streamers and platforms \cite{mt19infocom}. HCI communities investigate the reasoning of new user behavior (e.g., donation) in live streaming \cite{donate18CHI,donate18china}. However, they are not designed for recommendations. In this paper, we make the first attempt to develop donation and \mwp\ recommenders. 

\noindent\textbf{Personalized Recommendation.} 
Personalized recommender systems learn user preferences of items from user feedback \cite{XH17,zhang18lkg}. Collaborative Filtering (CF) \cite{cse19www,XH17} leverages auxiliary information, such as item similarities, to derive patterns for recommendations. Knowledge Graph-based methods \cite{zhang18lkg} learn low-dimensional embeddings of users and items from heterogeneous and structural information. However, these methods do not encourage social interactions since they only focus on recommending items for individuals, not to mention the viewer-streamer interactions (e.g., donations). Existing online video platforms (e.g., YouTube) analyze the contents and contexts of videos for recommendation but do not take donations into consideration \cite{ChenZ0NLC17,LiuCLH19}. 

\noindent\textbf{Group Recommendation}.
Group recommenders aggregate features among users in a group as group features \cite{vldb20vr,DC18}. Cao et al. learn user influence in a group with an attention network to infer group consensus \cite{DC18}. Shen et al. \cite{CYS18} select a group of socially-tight users and a set of preferred channels jointly to foster social interactions. However, the channels are identical for each individual, whereas their diverse personal interests are not considered. Although the trade-off between personal and group interests in VR shopping are addressed \cite{cikm19vr,vldb20vr}, they support neither the donation recommendations nor streamer relations.

\begin{figure}[tp]
  \includegraphics[width = 0.95\columnwidth]{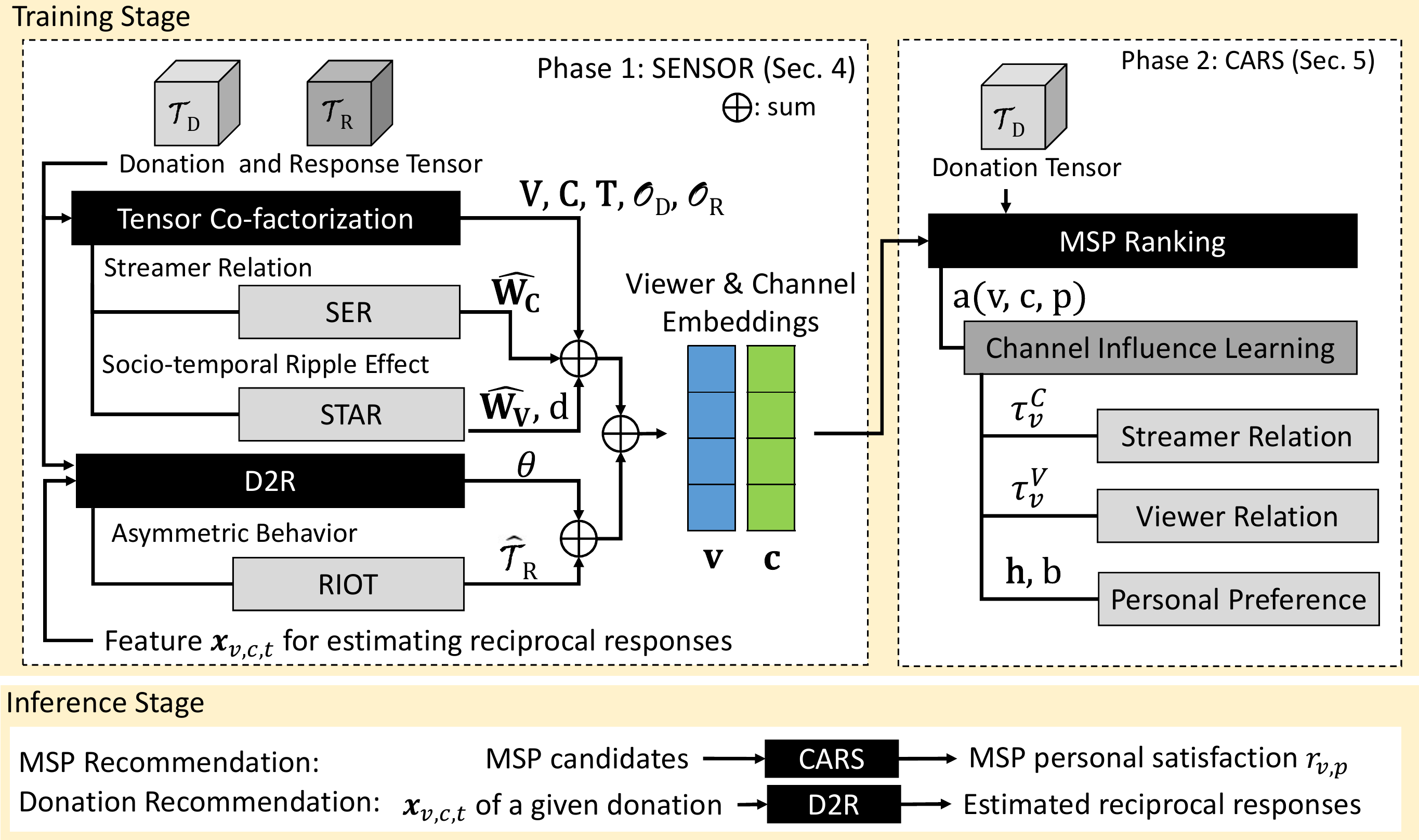}
  \caption{The proposed framework \framework.}
  \label{fig:framework}
\end{figure}

\section{Overview of \prob}\label{sec:method}
To facilitate a satisfactory multi-stream party, two key tasks are to 1) recommend donations for viewers such that the streamers give the maximum reciprocal response, and 2) rank the \mwp\ personal satisfaction of candidate \mwp s for each viewer to facilitate \mwp\ recommendations. Figure \ref{fig:framework} illustrates the two-phase framework \framework\ proposed to support these two tasks. The yellow rectangle at the top shows the training stage of \framework. In phase one (dotted rectangle on the left), \tensor\ has two goals (black rectangles). The first task is viewer and channel embedding extraction, which is achieved by the proposed tensor co-factorization technique. Moreover, \textit{\underline{S}treamer R\underline{e}lation \underline{R}egularization (\ser)} and \textit{\underline{S}ocio-\underline{T}emporal \underline{A}utoregressive \underline{R}egularization (\ripple)} (grey rectangles) are proposed to regularize the impact of streamer relations and the socio-temporal ripple effect in live streaming. The second task is \textit{\underline{D}onation \underline{to} \underline{R}esponse estimation (\dtor)}, which estimates the reciprocal responses of a given donation for donation recommendations. In addition, we propose \textit{\underline{R}esponse Suppress\underline{i}on in D\underline{o}nation Burs\underline{t} (\smooth)} to model with the asymmetric behaviors of viewers and streamers during a donation burst. As such, \dtor\ avoids recommending donations with low-quality reciprocal responses. In the second phase (dotted rectangle on the right), the goal of \rank\ is to learn an \mwp\ ranking function (black rectangle) based on the viewer and channel embeddings derived in \tensor, and the channel influence (dark grey rectangle), comprised of streamer relation, viewer relation, and personal interest aspects (grey rectangles below). Thus, \rank\ considers the interplay in different combinations of channels and friends in an \mwp\ when quantifying \mwp\ personal satisfaction. 

In the inference stage (yellow rectangle below) for \mwp\ recommendations, given a set of candidate \mwp s, we first derive each viewer's \mwp\ personal satisfaction with \rank. The widely-used Least-Misery (LM) mechanism can be adopted to recommend the best \mwp\ with the greatest minimum \mwp\ personal satisfaction among the viewers. While watching \mwp, the system shows the estimated response based on \dtor\ for each channel. Hence, the viewers can send the donation amount along with a message to the channel with the maximum estimated reciprocal responses.

\subsection{Research Challenges}\label{sec:challenge}
\noindent\textbf{C1: The seesaw in optimizing distinct goals.} First, learning to recommend an \mwp\ and to recommend donations jointly could lead to underfitting due to their different goals. Moreover, it is important to carefully examine the correlations between them. In other words, it is necessary to consider the potential donation recommendations before recommending an \mwp, such that the recommended \mwp\ not only satisfies the viewers but also allows them to receive the reciprocal responses from the streamers easily. In \framework, \tensor\ extracts the embeddings with a novel tensor co-factorization model to profile the social and temporal behaviors of viewers and channels in the first phase. Subsequently, \rank\ ranks candidate \mwp s \textit{based on the co-factorized embeddings} to correlate two goals. 

\noindent\textbf{C2: Complicated streamer relations and viewer relations.} Donation recommendation targeting maximum reciprocal responses is challenging due to unique social phenomena in live streaming platforms. \textit{Streamer relations} affect viewers' donations e.g., some viewers prefer watching the debate between rivals and make donations to support them. Moreover, viewers are more willing to donate after other viewers are also donating \cite{donate18CHI,donate18china}, termed as \textit{the socio-temporal ripple effect of donation}, which may be attributed to competition for attention. Although some existing temporal social-aware recommenders \cite{sbpr16cikm} can be used for donation recommendations, they do not model the relationships between streamers. In contrast, \tensor\ includes two regularizers, namely \ser\ and \ripple, to derive the signed streamer relations and the social influence among other viewers with influence decay, respectively.

\noindent\textbf{C3: Asymmetric viewer and streamer behaviors.} The streamers are usually distracted during a burst of donations since they face many viewers \textit{alone}, whereas the viewers only watch a few channels. This asymmetric communication between a streamer and his viewers may lower viewers' satisfaction with his donation due to low-quality responses. Existing temporal feedback prediction methods \cite{time19kdd,time17icdm} do not model this unique phenomenon. In \dtor, \smooth\ is introduced to model the distraction behavior during the burst so \dtor\ can effectively avoid recommending donations while a burst is happening.

\noindent\textbf{C4: Tradeoff between personal interests and group interactions.} To recommend \mwp s, it is challenging to strike a balance between the tradeoff in various \mwp s to quantify the \mwp\ personal satisfaction for recommendations. One possible approach is to employ personalized recommendations \cite{XH17,cse19www} for each individual. However, the complicated social effects occurring amongst viewers watching common channels are not carefully examined. On the other hand, existing social-aware recommendations \cite{sbpr16cikm,rmse18ijcai} only infer personal satisfaction based on social topology, which is not designed for \mwp\ where viewers watch live streaming channels together. Hence, we parameterize \mwp\ personal satisfaction from personal, social, and streamer relation aspects, to design a new ranking model \rank\ to quantify the satisfaction of different combinations of channels and friends.

\subsection{Preliminaries}\label{sec:preliminaries}
For clarity of presentation, in this paper, bold uppercase letters (e.g., $\mathbf{X}$) and lowercase letters (e.g., $\mathbf{x}$) denote matrices and column vectors, respectively.  $\mathbf{X}(i,j)$ and $\mathbf{X}(i,:)$ denote the element in the $i$-th row of the $j$-th column and the $i$-th row vector, respectively. Non-bold letters (e.g., $x$) and squiggle letters (e.g., $\mathcal{X}$) represent scalars and tensors, respectively. Let $G=(V,E)$ be a social network of viewers, $C$ be a channel set, $G_{MW}=\{g_1,\cdots,g_n,\cdots,g_N\}$ be a \mwtech\ group set, where $g_n=\{V_n, E_n\}$ is an induced subgraph of $G$, and the set of their corresponding \mwpfull\ (\mwp) is $P_{MW}=\{p_1,\cdots,p_n,\cdots,p_N\}$, where $p_n=\{C_{n,v}|v\in V_n,C_{n,v}\subseteq C^k\}$ consists of a set $C_{n,v}$ of $k$ channels for each viewer $v$ in $V_n$ to watch. On the other hand, $\mathbf{W}_C\in \{-1,0,+1\}^{|C|\times |C|}$ denotes the signed network of streamers, where $\mathbf{W}_C(i,j)=1$ if streamers $i$ and $j$ have a positive relation (e.g., friends), $-1$ if they have a negative relation (e.g., rivals), and $0$ if they don't have a relation. In addition, the donations and the streamer responses are respectively recorded in the donation tensor $\mathcal{T}_D$ and the response tensor $\mathcal{T}_R$\footnote{Streamers provide various responses, e.g., gifts or verbal sentences, to viewers for expressing their gratitude. When the data lack of ground truth of the quality of reciprocal responses, a possible solution is to learn from multiple feedback (e.g., the length of verbal responses, the sentiments of verbal responses, and the number of gifts sent) from streamers.}, where $\mathcal{T}_D, \mathcal{T}_R\in\mathbb{R}^{|V|\times |C|\times T}$ and $T$ is the number of time slots. An element $\mathcal{T}_D(v,c,t)$ indicates how much money viewer $v$ donates to channel $c$ at timestamp $t$, while $\mathcal{T}_R(v,c,t)$ denotes the measured quality of the corresponding reciprocal responses.

Figure \ref{fig:mwp1} illustrates an example of \mwp\ ($p_n$) with $k=3$, i.e., every viewer watches three channels, for a group $g_n$ where $V_n=\{v_1,v_2,v_3,v_4\}$ and $E_n=\{(v_1,v_2),(v_2,v_3),(v_3,v_4),(v_2,v_4)\}$. $\mathbf{W}_C(c_S,c_D)=-1$ denotes that channels $c_S$ and $c_D$ have a negative relationship. $C_{n,v_2}=\{c_S,c_E,c_D\}$ denotes that $v_2$ watches channels $c_S$, $c_E$, and $c_D$. Finally, the \probfull~(\prob) problem is formally defined as follows.

\noindent\textbf{Problem: \probfull~(\prob).}

\noindent\textbf{Given:} $\mathcal{T}_D$, $\mathcal{T}_R$, $G$, $G_{MW}$, $P_{MW}$, $k$, and $\mathbf{W}_C$.

\noindent\textbf{Goal 1:} Estimate the expected reciprocal response of a given donation (Section \ref{sec:tensor}).

\noindent\textbf{Goal 2:} Rank candidate \mwp s, including channels and group members (Section \ref{sec:rank}).

\noindent We respectively detail phases 1 and 2 of \framework~in Sections 4 and 5.

\section{Socio-temporal Donation-Response Tensor Co-factorization (SENSOR)}\label{sec:tensor}
Since the behaviors of donations and responses are highly related to each other, to extract embeddings of viewers and channels, we propose an innovative co-factorization model \tensor\ on $\mathcal{T}_D$ and $\mathcal{T}_R$ that shares low-dimensional latent spaces of viewers and channels. To tackle challenge C2, \ser\ factorizes \textit{the signed social networks of streamers} $\mathbf{W}_C$ to encode the streamer relations in the embeddings of streamers. Furthermore, to model the \textit{socio-temporal ripple effect of donations} into \tensor, we introduce \ripple\ to learn both a social influence matrix of viewers on donations $\hat{\mathbf{W}}_V$ with the degree of exponential decay of the influence $d$. Therefore, the total social influence of donations, which adds up the social influence of each donation weighted by time, becomes greater only if plenty of donations from influential viewers and friends happen within a recent and short period. \ripple\ penalizes the objective if the estimated user donation and its total social influence of donations are inconsistent.

On donation recommendations, an idea is to accurately estimate the reciprocal response from a streamer to a given donation and then recommend the viewer donating to the streamer who is expected to return the maximum reciprocal response. Therefore, we extract important features (viewer-streamer relations, message sentiment and semantics, etc) to learn a model \dtor. Moreover, to tackle C3, we propose \smooth, which disperses the values of the reciprocal responses when the donations are made in a burst. Finally, as we can precisely estimate the reciprocal responses based on viewers' input donations and messages and avoid donation bursts, we can recommend the viewers donating to the streamers with the maximum estimated reciprocal response in real-time. \dtor\ and \smooth\ are integrated into \tensor\ rather than learned alone since they also help interpret the relations between $\mathcal{T}_D$ and $\mathcal{T}_R$ during factorization.

Our final goal is to extract the latent features of each viewer and channel via modeling the socio-temporal effect of donations and responses for ranking \mwp s. As the viewer donations and streamer responses are naturally represented as tensors, we design a new coupled tensor factorization model for jointly learning the viewer and channel embeddings to rank \mwp s. Specifically, to capture the essential properties of donations and responses, we collectively factorize $\mathcal{T}_D$ and $\mathcal{T}_R$ into unified low-dimensional representation matrices by minimizing the following function:
\begin{align*}
    \|\mathcal{T}_D-\mathcal{O}_D\times_1\mathbf{V}\times_2\mathbf{C}\times_3\mathbf{T}\|^2+
    \|\mathcal{T}_R-\mathcal{O}_R\times_1\mathbf{V}\times_2\mathbf{C}\times_3\mathbf{T}\|^2,
\end{align*}
where $\mathbf{V}\in\mathbb{R}^{|V|\times\alpha}$, $\mathbf{C}\in\mathbb{R}^{|C|\times\alpha}$, and $\mathbf{T}\in\mathbb{R}^{T\times\alpha}$ are the trainable matrices with respect to viewers, channels, and timestamps. $\mathcal{O}_D,\mathcal{O}_R\in \mathbb{R}^{\alpha^3}$ are the trainable core tensors with $\alpha$ representing the dimensionality of latent features. $\|\mathcal{T}\|=\sqrt{\langle \mathcal{T},\mathcal{T}\rangle}$ denotes the Frobenius norm of a tensor $\mathcal{T}$, and $\times_t$ denotes the $t$-mode product. 

In the above formulation, $\mathbf{V}$, $\mathbf{C}$, and $\mathbf{T}$ are shared between the factorization of $\mathcal{T}_D$ and $\mathcal{T}_R$ to jointly construct the latent representations, while the core tensors, i.e., $\mathcal{O}_D$ and $\mathcal{O}_R$, are respectively used to model the interactions between viewers, channels and timestamps in donations and responses. An intuition is to factorize $\mathcal{T}_D$ and $\mathcal{T}_R$ separately with Tucker Decomposition \cite{Tucker} and concatenates them together to obtain $\mathbf{V}$, $\mathbf{C}$, and $\mathbf{T}$. However, it is not efficient according to the following proposition.
\begin{proposition}
Sharing common latent matrices reduces twice of the amounts of parameters than separately factorizing them with Tucker Decomposition \cite{Tucker}.
\end{proposition}
By sharing common latent matrices, it takes $(|V|+|C|+T)\alpha+2\alpha^3$ parameters. In contrast, in the separated case, $\mathcal{T}_D$ and $\mathcal{T}_R$ independently generate $\mathbf{V}$, $\mathbf{C}$, and $\mathbf{T}$, which needs $2(|V|+|C|+T)\alpha+2\alpha^3$ parameters in total. Therefore, the amount of parameters is almost half ($\alpha$ is usually small) for the proposed approach without duplicating latent matrices, which reduces enormous parameters to avoid overfitting and accelerate the computation time. Moreover, sharing common latent matrices preserves the correlations of viewers, channels, and timestamps in the latent space.

Another possible approach is to use a 4-dimensional tensor, where the 4th dimension indicates donation and response. However, it is not efficient due to the following proposition.
\begin{proposition}
The proposed model reduces the amount of parameters from $(|V|+|C|+T+2)\alpha+\alpha^4$ (in a 4-dimensional tensor) to $(|V|+|C|+T)\alpha+2\alpha^3$, and reduces the computation of gradients by half.
\end{proposition}
That is, besides $\mathbf{V}$, $\mathbf{C}$, and $\mathbf{T}$, factorizing the 4-dimensional tensor will incorporate an additional matrix, namely $\mathbf{A}$, with $2\times\alpha$ entries, storing the latent features of actions of donation and response. The core tensor will be extended to $\alpha^4$ for a 4-mode product. Therefore, the space complexity for the 4-dimensional tensor, which is $(|V|+|C|+T+2)\alpha+\alpha^4$, is greater, which may cause overfitting. On the other hand, computing the gradients of $\mathbf{V}$, $\mathbf{C}$, and $\mathbf{T}$ in each SGD iteration requires an additional 4-mode product with the vector $\mathbf{a}$, which is a column vector of $\mathbf{A}$, with 2 entries, leading to double computation time. Finally, the loss function of the proposed \tensor, namely $\mathcal{L}_{\tensor}$, is derived as follows:
\sloppy 
\begin{align*}
    \mathcal{L}&_{\tensor}=\mathcal{L}_{D2R}+\lambda_1R_{\ser}+\lambda_2R_{\ripple}+\lambda_3R_{\smooth}\\
    &+\|\mathcal{T}_D-\mathcal{O}_D\times_1\mathbf{V}\times_2\mathbf{C}\times_3\mathbf{T}\|^2+\|\mathcal{T}_R-\mathcal{O}_R\times_1\mathbf{V}\times_2\mathbf{C}\times_3\mathbf{T}\|^2,
\end{align*}
where $R_{\ser}$, $R_{\ripple}$, and $R_{\smooth}$ are regularizers for \ser, \ripple, and \smooth, respectively, while $\mathcal{L}_{D2R}$ is the loss function of reciprocal response estimation (detailed right next). $\lambda_1$, $\lambda_2$, and $\lambda_3$ control the weights of each regularizer. For simplicity, $\mathbf{v}=\mathbf{V}^\intercal(v,:)$ and $\mathbf{c}=\mathbf{C}^\intercal(c,:)$ respectively denote the viewer embedding of viewer $v$ and the channel embedding of channel $c$. $\hat{\mathcal{T}_D}=\mathcal{O}_D\times_1\mathbf{V}\times_2\mathbf{C}\times_3\mathbf{T}$ and $\hat{\mathcal{T}_R}=\mathcal{O}_R\times_1\mathbf{V}\times_2\mathbf{C}\times_3\mathbf{T}$ are the estimated donation and response tensors, respectively.

\noindent\textbf{SER: Streamer Relation Regularization.}
Viewers enjoy watching streamers interacting with each other. As shown in the real example in Figure \ref{fig:squad}, the competition between streamers Steven and Dan attracts 3K viewers watching and donating. Hence, it is crucial to encode \textit{streamers' relationships} in streamer embeddings. We design the \ser\ by factorizing $\mathbf{W}_C$ as follows:
\begin{align*}
    R_{\ser}=\|\mathbf{W}_C-\hat{\mathbf{W}}_C\|^2=\|\mathbf{W}_C-\mathbf{C}^\intercal\mathbf{C}\|^2,
\end{align*}
where each element in $\hat{\mathbf{W}}_C=\mathbf{C}^\intercal\mathbf{C}$ denotes the derived streamer relation between each pair of streamers. By minimizing the loss between $\mathbf{W}_C$ and $\hat{\mathbf{W}}_C$, the embeddings $\mathbf{c}$ and $\mathbf{\tilde{c}}$ of two streamers $c$ and $\tilde{c}$ with a positive relationship are inclined to be close in the latent space (i.e., $\mathbf{c}^\intercal\mathbf{\tilde{c}}$ is close to 1). Otherwise, they are put farther in the latent space (i.e., $\mathbf{c}^\intercal\mathbf{\tilde{c}}$ is close to -1).

\noindent\textbf{STAR: Socio-Temporal Autoregressive Regularization.}
Previous works show that a viewer is more willing to donate after seeing others' donations, especially friends~\cite{donate18CHI,donate18china}. In other words, the donations have a \textit{socio-temporal ripple effect} triggered by other donations. From our collected Twitch dataset, the probability that a donation follows another donation within 5 minutes is 30.4\%. Research \cite{donate18CHI} also manifests that viewer donations to streamers may draw other viewers' attention, which causes further interactions (i.e., herd behavior), including donations. Therefore, we design a novel regularization term $R_{\ripple}$ as follows:
\begin{align}
    R_{\ripple}&=\sum_{\forall v,c,t}(\hat{\mathcal{T}}_D(v,c,t)\notag\\
    &-\sum_{u\in V-\{v\}}\sum_{\Delta_t=1}^{L}[e^{-d\Delta_t}\hat{\mathbf{W}}_V(u,v)\mathcal{T}_D(u,c,t-\Delta_t)])^2,\label{eq:star}
\end{align}
where $\hat{\mathbf{W}}_V\in\mathbb{R}^{|V|\times|V|}$ is the learnable weight matrix of social influence between viewers. $e^{-d\Delta_t}$ measures the exponential decay in previous $L$ time slots, where $\Delta_t\in[1,L]$ is the temporal difference and $d$ is the decay factor. Hence, the socio-temporal impact of the donation $\mathcal{T}_D(u,c,t-\Delta_t)$ to viewer $v$ is stronger if the donation time is closer ($\Delta_t$ is smaller) and if the donor $u$ has a high influence on $v$ ($\hat{\mathbf{W}}_V(u,v)$ is greater). We initialize $\hat{\mathbf{W}}_V(u,v)=1$ if and only if $(u,v)\in E$; and $\hat{\mathbf{W}}_V(u,v)=\epsilon$ otherwise. Moreover, different friends (e.g., close friends vs. common friends) or different non-friend viewers (e.g., celebrities vs. common viewers) may have different influences. As in Figure \ref{fig:framework}, \ripple\ learns $\hat{\mathbf{W}}_V$ and $d$ from data to restrict $\hat{\mathcal{T}}_D(v,c,t)$ to be close to the estimation from previous donations by Eq. (\ref{eq:star}). Hence, both \ser\ and \ripple\ help embed streamer and viewer behaviors while factorizing their embeddings.

\noindent\textbf{D2R: Donation-to-Response Estimation.}
Streamers reply to donations to interact with viewers and express their gratitude. Research shows that their responses strengthen the engagement of viewers to the channels \cite{donate18CHI,donate18china}. Nevertheless, viewers usually have limited budgets for donations. Therefore, it is necessary to estimate the reciprocal response potentially received for a given donation in donation recommendations for viewers. Given a donation $\mathcal{T}_D(v,c,t)>0$, we extract the following features as the input of estimating the expected reciprocal response: 1) the amount of the donation $\mathcal{T}_D(v,c,t)$, 2) the sentence embeddings of the donation messages \cite{TM13NIPS}, 3) the text sentiment scores of the donation messages \cite{sentiment16}, 4) the real-time speech emotions and sentiments of $c$ \cite{emotion16emnlp}, 5) the cumulative amount of donations of $c$ from $v$, 6) the cumulative amount of donations of $c$ in the recent $L$ slots, 7) the minimum donation amounts to be shown on the top fan list, and 8) the viewer-streamer interactions $\mathbf{v}\odot\mathbf{c}$ ($\odot$ is the element-wise product). The above features are concatenated into a feature vector $\mathbf{x}_{v,c,t}$. Accordingly, the loss for \dtor\ $\mathcal{L}_{D2R}$ is defined as follows:
\begin{align*}
    \mathcal{L}_{D2R}=\sum_{\forall v,c,t\land\mathcal{T}_D(v,c,t)\neq 0}(\hat{\mathcal{T}}_R(v,c,t)-\theta^\intercal\mathbf{x}_{v,c,t})^2,
\end{align*}
where $\theta$ is the weight vector regressing $\mathbf{x}_{v,c,t}$ to the estimated reciprocal response $\hat{\mathcal{T}_R}(v,c,t)$. The square of difference is minimized as the auxiliary loss for factorization. In the inference stage (as shown in Figure \ref{fig:framework}), the streamer with the maximum estimated reciprocal response, derived by $\theta$ and $\mathbf{x}_{v,c,t}$, of a given donation is recommended to the viewer.

\noindent\textbf{RIOT: Response Suppression in Donation Burst.}
Unlike viewers who only focus on a few streamers, streamers usually interact with many viewers. As a result, they are usually distracted during a burst of donations since they broadcast to viewers \textit{alone} \cite{donate18CHI}. When plenty of donations occur in a short period, the streamer is physically constrained from a long engagement with individual donors and thus is inclined to respond less and shorter. To model this asymmetric communication phenomenon, we use a regularization term to help \dtor\ better estimate reciprocal responses. The goal of the regularization is to suppress the estimated reciprocal responses to a small value if the donations burst in a short period $L$. Therefore, for each channel $c$, we calculate the joint donation distribution of  $v$ and $t'$ as follows:
\begin{align*}
    p^D_{v,c,t'}&=\frac{\mathcal{T}_D(v,c,t')}{\sum_{v\in V,t'\in[t-L,\cdots,t]}\mathcal{T}_D(v,c,t')}.
\end{align*}
The entropy of the joint donation distribution in $[t-L,t]$, denoted as $S_D(c,L,t)$, is derived as follows.
\begin{align*}
    S_D(c,L,t)&=-\sum_{v\in V,t'\in[t-L,\cdots,t]}p^D_{v,c,t'}\log(p^D_{v,c,t'}).
\end{align*}
When the donation bursts, $p^D_{v,c,t'}$ is likely to be small within $L$ so the entropy $S_D(c,L,t)$ increases. Accordingly, we design a regularization term $R_{\smooth}$ to suppress the values in the derived response tensor $\hat{\mathcal{T}_R}$ based on the entropy $S_D(c,L,t)$.
\begin{align*}
    R_{\smooth}&=\sum_{c\in C, t\in T}\phi_{c,t}\cdot S_D(c,L,t)\cdot\hat{\mathcal{T}_R}(v,c,t'),
\end{align*}
where $\phi_{c,t}=\sum_{v\in V}\mathcal{T}_D(v,c,t)-\frac{\sum_{v\in V,t'\in[t-L,\cdots,t]}\mathcal{T}_D(v,c,t')}{L+1}$ is the difference between the total amount of donations at time $t$ and the average amount of donations within $L$, which  measures the trend of donation bursts in channel $c$ at time $t$. In other words, $\phi_{c,t}$ is positive if the donation amount is increasing, and hence $\hat{\mathcal{T}_R}(v,c,t')$ would be lower. Otherwise, the donation amount is decreasing and the streamer is likely free to offer high-quality reciprocal responses.
Moreover, while $S_D(c,L,t)$ is large, i.e., the donation is bursting, $\hat{\mathcal{T}_R}(v,c,t')$ becomes small to reduce the loss. As a result, \smooth\ is able to capture the asymmetric communication phenomenon in a donation burst. Note that $\phi_{c,t}$ and $S_D(c,L,t)$ are fixed when minimizing $R_{\smooth}$. In summary, \tensor\ is designed to: 1) recommend donations with \dtor, and 2) derive latent matrices $\mathbf{V}$ and $\mathbf{C}$ for ranking \mwp s (the next section).

\section{Channel Influence-Aware MSP Ranking System (CARS)}\label{sec:rank}
The second phase of \framework\ aims to quantify the \mwp\ personal satisfaction of viewer $v$ on \mwp\ $p$, denoted as $r_{v,p}$, to facilitate \mwp\ recommendations. A na\"ive method to obtain the \mwp\ embedding of $p$, denoted as $\mathbf{p}$, is to use the weighted sum of channel embeddings and model the \mwp\ personal satisfaction as the inner product of $\mathbf{v}$ and $\mathbf{p}$~\cite{ChenZ0NLC17,geofm15sigir}, i.e., $r_{v,p}=\mathbf{v}^\intercal\mathbf{p}$. However, it is not practical since a viewer may pay more attention to those channels she likes, those channels she shares with friends, or those streamers who interact with each other. To address \textit{the tradeoff between personal interests and group interactions} (C4) as well as the streamer relations, we propose to parameterize the \textit{channel influence} for each viewer-channel pair. By factoring the personal, social, and streamer relation aspects in channel influence, we respectively learn the weights $\tau^V_v$ and $\tau^C_v$ corresponding to the social part and the streamer relation part for a viewer $v$. With the channel influence, we redefine the \mwp\ personal satisfaction and propose a new ranking model \rank\ to minimize the total pairwise ranking loss based on the total amounts of donations made in an \mwp\ by a viewer. 

\noindent\textbf{Learning Channel Influence.} Specifically, we parameterize the influence $a(v,c,p)$ of a viewer-channel pair $(v,c)$ conditioned on an \mwp\ $p$ as follows:
\begin{align*}
    o(v,c)&=\mathbf{h}^\intercal\sigma(\mathbf{v}\oplus\mathbf{c}\oplus b)\\
    a(v,c,p)&=o(v,c)+\tau^{V}_v\sum_{(u,v)\in E_p\land c\in C_{p,u}}\hat{\mathbf{W}}_V(u,v)\cdot o(u,c)\\
    &+\tau^{C}_v\sum_{\tilde{c}\in C_{p,v}\land c\neq \tilde{c}}|\mathbf{c}^\intercal\mathbf{\tilde{c}}|,
\end{align*}
where $\oplus$ denotes the vector concatenation and $b$ is the bias. A sigmoid function $\sigma$ is used as a gating function to identify important features and project those important features to the original influence $o(v,c)$ of $c$ for $v$ with a vector $\mathbf{h}$. Note that the behavioral features of viewer donation and streamer reciprocal responses are conducted by using $\mathbf{v}$ and $\mathbf{c}$ here. $a(v,c,p)$ comprises a personal part $o(v,c)$, a social part $\sum_{(u,v)\in E_p\land c\in C_{p,u}}\hat{\mathbf{W}}_V(u,v)\cdot o(u,c)$, and a streamer relation part $\sum_{\tilde{c}\in C_{p,v}\land c\neq \tilde{c}}|\mathbf{c}^\intercal\mathbf{\tilde{c}}|$, where $u$, $v$ are friends and they share the common channel $c$ together in $p$. For the social part, $\hat{\mathbf{W}}_V$ is the social influence matrix derived from \tensor\ and $\tau^{V}_v$ is the customized contribution factor of the social part for each viewer $v$. Note that learning $o(v,c)$ and $\tau^V_v$ allows us to obtain the composition of the preferences of a viewer regarding an \mwp. For instance, if $v$ prefers watching with friends, $\tau^V_v$ is large, enhancing the importance of the social part, leading to satisfactory watching experience and potential donations.

Besides, if the streamers have stronger relations (e.g., teammates and opponents), their interactions (e.g., collaborations or competitions) may also attract viewers' attention. We sum up the absolute value $|\mathbf{c}^\intercal\mathbf{\tilde{c}}|$ of the derived streamer relation between streamers $c$ and $\tilde{c}$ to represent the effectiveness in channel influence due to streamer relations. Here the absolute value is adopted since positive and negative relations are both inclined to create intensive interactions. $\tau^{C}_v$ is the customized contribution in channel influence of streamer relations for each viewer $v$. If $v$ likes streamer interactions, $\tau^C_v$ becomes larger. Thus, recommending streamers with stronger relations to $v$ may incur more donations.

Finally, $a(v,c,p)$ is larger if 1) $v$ likes $c$ a lot (large $o(v,c)$), 2) her close friend $u$ (large $\hat{\mathbf{W}}_V(u,v)$) also likes $c$ a lot (large $o(u,c)$), or 3) the streamers in $C_{p,v}$ have strong relations (large $|\mathbf{c}^\intercal\mathbf{\tilde{c}}|$). We redefine \mwp\ personal satisfaction $r_{v,p}$ as follows:
\begin{align*}
    r_{v,p}&=\mathbf{v}^\intercal\sum_{c\in C_{p,v}} a(v,c,p)\cdot \mathbf{c},
\end{align*}
where the \mwp\ personal satisfaction is the inner product of the viewer embedding of $v$ and the sum of the embeddings of channels watched by $v$ in $p$, weighted by $a(v,c,p)$. By learning $a(v,c,p)$, \rank\ analyzes a viewer's satisfaction from personal, social, and streamer relation aspects. However, personalized recommenders fail to consider the social aspect \cite{XH17,cse19www}, while group \cite{DC18,vldb20vr} and package \cite{CYS18} recommenders do not consider the streamer relations.

\sloppy
\noindent\textbf{\mwp\ Ranking.} Viewers seldom provide their explicit feedback (e.g., ratings) after watching live streaming channels. Therefore, inspired by BPR \cite{SR09}, we learn to rank the \mwp s by comparing pairs of distinct \mwp s. First, we construct a training dataset of BPR as $DB=\{(v,p,p')|\sum_{\forall c\in p}\sum_{\forall t\in T}\mathcal{T}_D(v,c,t)>\sum_{\forall c'\in p'}\sum_{\forall t\in T}\mathcal{T}_D(v,c',t)\land v\in V\land p,p'\in P_{MW}\}$. That is, $v$ prefers $p$ to $p'$ if she has donated more to $p$ than to $p'$. A variant is to compare \mwp s based on the amounts of the received reciprocal responses (i.e., $\sum_{\forall c\in p}\sum_{\forall t\in T}\mathcal{T}_R(v,c,t)$) since viewers donate to seek social interactions \cite{donate18CHI,donate18china}. However, it may lead to a biased training result that prioritizes unpopular channels since those streamers are free to respond more \cite{donate18CHI}. In contrast, optimizing the rank of donations potentially improves the social engagement of viewers since they enjoy the recommendation and donate more, which also leads to more profits for streamers and the platforms.

Equipped with $a(v,c,p)$, we propose a novel ranking system \rank\ with a new BPR-based ranking loss $\mathcal{L}_{\rank}$ as follows:
\begin{align*}
  \mathcal{L}_{\rank}(\Theta)=\sum_{(v,p,p')\in DB}-\ln\sigma(r_{v,p,p'})+\frac{\lambda_4}{2}\|\Theta\|^2_2,
\end{align*}
where $r_{v,p,p'}=r_{v,p}-r_{v,p'}$ is the difference of ranking scores for a paired \mwp\ instance $(v,p,p')\in DB$, $\sigma(\cdot)$ maps $r_{v,p,p'}$ to a value between 0 and 1, $\ln(\cdot)$ is the log-likelihood, and $\lambda_4$ controls the impact of the sparsity regularizer. $\Theta=\{\mathbf{h},b\}\cup\{\tau^V_v,\tau^C_v|\forall v\in V\}$ is the set of model parameters for learning $a(v,c,p)$ (as in Figure \ref{fig:framework}). We do not vary viewer and channel embeddings here since it may deteriorate the estimation of the reciprocal responses in Section \ref{sec:tensor}. 

\section{Experimental Results}\label{sec:exp}
\subsection{Experiment Setting}
\subsubsection{Baselines}\label{apdx:baseline}
\noindent\textbf{Donation recommendations.}
We compare \dtor\ with \dtorn\ (\dtor\ without \smooth) and state-of-the-art feedback prediction methods: \textit{ATS} \cite{time19kdd}, \textit{STNN} \cite{time17icdm}, and \textit{DLR} \cite{linearregression}. ATS employs an attention network to identify important features for time series prediction. STNN introduces a dynamic RNN for time series prediction in spatio-temporal cases. The weighted adjacency matrix of POIs in STNN is replaced by the streamer relations to consider socio-temporal behaviors for a fair comparison. DLR is a regression model on a series of observed donation amounts and the corresponding reciprocal responses. We compare \tensor\ with Tucker Decomposition \cite{Tucker} and conduct an ablation study to examine the effectiveness of each term in \tensor, i.e., \textit{\tensorser} (\tensor\ without \ser), \textit{\tensors} (without \ripple), \textit{\tensorr}(without \smooth), and \textit{\tensorn}  (without all of the regularization terms).

\noindent\textbf{\mwp\ recommendations.} 
We compare \rank\ with the state-of-the-art methods: personalized recommender \textit{NCF} \cite{XH17}, social-aware personalized recommender \textit{SBPR}~\cite{sbpr16cikm}, group recommender \textit{GBPR}~\cite{WP13IJCAI}, and hybrid recommender \textit{AGREE}~\cite{DC18}. NCF ranks the personal interests with a deep neural network. SBPR learns viewer influence and ranks items jointly. GBPR employs Matrix Factorization to find group consensus without considering channel influence. AGREE jointly learns group consensus and personal interests with an attention network. We also compare \rank\ with its variants: 1) \textit{\ranku} with unified channel influence (i.e., $a(v,c,p)=\frac{1}{k},\forall v,c,p$), 2) \textit{\rankf} with binary friendships (i.e., $\hat{\mathbf{W}}_V\in\{0,1\}^{|V|\times |V|}$), 3) \textit{\rankc} without considering streamer relations ($\tau^C_v=0,\forall v$), and 4) \textit{\rankn} without \tensor, which trains the embeddings from end-to-end.

\subsubsection{Dataset and pre-processing.}
\noindent 1) \textbf{Twitch-Full} is a live streaming dataset with 600K viewers, 140 channels, and 86K donations. We choose 35 channels with the greatest viewership for each content category, e.g., Just Chatting, League of Legends (LOL), etc. Note that Streamers of Just Chatting chat with viewers rather than broadcast gaming content. Since the viewers did not label their reciprocal responses online, we hire 43 workers, who are heavy users spending at least 2.5 hours per day on Twitch with extensive donation experiences, to watch the recorded live streaming videos (broadcasted in 2019) and manually label 1) the donation time and amounts and 2) the quality of reciprocal responses. Multiple measurements of reciprocal responses, i.e., the length of verbal sentences, sentiment scores, and response time (from 1: very slow to 5: very fast), are labeled on the Likert scale. The maximum of a given response labeled by a worker is selected as the worker's answer. Finally, the labels from at least three workers are averaged for each reciprocal response. The streamer relations are labeled manually as positive if they are teammates or friends, and negative if they are rivals. The viewer social network is crawled via Twitch APIs.

\noindent 2) \textbf{Twitch-Chat} is a chat log of 2,162 streaming from 52 channels and 2.04M viewers \cite{twitch_chat_data}. Viewers leave comments in public chat rooms (open to all the viewers and the streamer in a channel) in order to interact with streamers and other viewers, similar to donation behaviors~\cite{donate18CHI,donate18china}. As a result, we simulate the chats as donations in binary format, i.e., $\mathcal{T}_D(v,c,t)=1$ if viewer $v$ leaves at least one comment in channel $c$ at timestamp $t$; and $\mathcal{T}_D(v,c,t)=0$ otherwise. $\hat{\mathbf{W}_V}$ is initialized with small random values following the settings in \cite{scl15random}, since the viewer social network is not given. $\mathbf{W}_C$ and reciprocal responses are not provided. Therefore, we only examine \mwp\ recommendations but not donation recommendations.

\noindent 3) \textbf{Douyu} is collected with the donations (in terms of virtual gifts) from the live streaming platform Douyu \cite{douyu_data}, including 242K channels, 7M viewers, and 64.9M donations. The settings of $\hat{\mathbf{W}_V}$, $\mathbf{W}_C$ and reciprocal responses are identical to Twitch-Chat. We examine \mwp\ recommendations but not donation recommendations, either.

\noindent 4) \textbf{Reddit} is a social bulletin board dataset, where users can post comments and reply to the posts with subreddits organized under the same threads, with more than 500M users and 1.7 billion public comments.\footnote{\url{https://www.kaggle.com/reddit/reddit-comments-may-2015}} Although Reddit is not a live streaming platform, it is public (unlike Facebook and Twitter) so posts are accessible to a large population (like live streaming). Users may flock to reply to the posts for discussions (socio-temporal ripple effect) but authors usually reply to only some of them (asymmetric communication behavior). To simulate the channels and viewers, following \cite{twitter10www,huang2018will}, we first extract the top 50K influential authors as the streamers by PageRank \cite{page1999pagerank}. The viewers are those who post subreddits below the streamers' comments. Moreover, to simulate an \mwp, we take the comments with the subreddits of a viewer within a short period (e.g., one hour) as the channels watched by the viewer. Two users' subreddits on the same comment during the same period represent that they view common channels. The subreddits are further treated as the donations, and the corresponding author replies as the responses. We remove stop words with NLTK \cite{nltk02} for every comment and subreddit to preserve meaningful words. The word counts in a subreddit and a comment respectively represent the amount of the donation and reciprocal response. $\hat{\mathbf{W}_V}$ is initialized with small random values \cite{scl15random}. By following \cite{kc17signednetwork,sk16signednetwork}, the elements in $\mathbf{W}_C$ are set to -1 if their conversations have negative words; 1 if they have a conversation without negative words; 0 otherwise.

For the ground truth of donation recommendations, ATS, STNN, and DLR use the labeled responses in Twitch-Full and the length of replies from the authors in Reddit. For the ground truth of \mwp\ recommendations, SBPR, GBPR, and NCF use the amounts of donations in Twitch-Full, Twitch-Chat, and Douyu, and the word counts of subreddits in Reddit as feedback. AGREE transforms the donations or ratings into binary labels (1 if a donation or rating is given; and 0 otherwise). In the inference stage, NCF and SBPR predict user satisfaction on a channel in an \mwp. GBPR and AGREE infer user satisfaction on a channel with group consensus if it is shared among friends, or with personal satisfaction otherwise. The \mwp\ personal satisfaction $r_{v,p}$ of an \mwp\ $p$ for a viewer $v$ in the above three approaches is obtained by the average of satisfaction upon the channels $v$ watched.

For each dataset, we extract 8K \mwp s and 10K viewers for training. For an \mwp , a group contains 5.3, 4.6, 6.3, and 9.7 users on average respectively in Twitch-Full, Twitch-Chat, Douyu, and Reddit, and each viewer watches $k=4$ channels. Existing platforms (e.g., Twitch, Mixer, NBC Sports, and Fox Sports) offer at most 4 channels simultaneously to a viewer to avoid overwhelming viewing experience. The hyper-parameters are optimized by five-fold cross-validation. Specifically, $\alpha =32$, $\epsilon =0.02$, $L=5$, $\lambda_2=\lambda_4=0.1$, $\lambda_1=\lambda_3=0.5$. Moreover, the length of a time slot is set as 2 minutes for Twitch-Full, Twitch-Chat, and Douyu, and 5 minutes for Reddit. 

\subsubsection{Evaluation Metrics}
In Section \ref{exp:donation}, the evaluation on donation recommendations is based on Root-Mean-Square Error (RMSE) \cite{time17icdm,rmse18ijcai}. For tensor factorization, we evaluate the average reconstruction loss with respect to donation and response tensors. In Section \ref{exp:mwp}, we measure the top-K Hit Ratio (HR@K) \cite{DC18} and the Mean Average Precision (MAP@K) \cite{DC18} to evaluate \mwp\ recommendations. We also show the efficiency in training datasets. We study the insights of \framework\ from Twitch-Full in Sections \ref{exp:social_influence} and \ref{exp:diff_channel}.

\begin{table}[t]
\footnotesize
\centering
\caption{Performances (RMSE) of different window size $L$ on donation recommendations.}
\label{donation_full}
\begin{tabular}{|c|c|c|c|c|c|c|}
\hline
       &           \multicolumn{3}{c|}{Twitch-Full}            &         \multicolumn{3}{c|}{Reddit}             \\ \cline{2-7} 
       &         $L=2$  &         $L=4$  &         $L=8$  &         $L=2$  &         $L=4$  &        $L=8$   \\\hline
DLR    &         6.03   &         5.87   &         7.22   &        12.97   &        10.99   &        13.41   \\
ATS    &         3.98   &         3.67   &         3.87   &         8.44   &         8.21   &         8.56   \\
STNN   &         3.61   &         3.57   &         4.00   &         8.22   &         8.07   &         8.24   \\\hline
\dtorn & \textbf{3.16}* & \textbf{2.79}* & \textbf{3.09}* & \textbf{7.20}* & \textbf{6.97}* & \textbf{7.11}* \\
\dtor  & \textbf{1.93}* & \textbf{1.72}* & \textbf{1.77}* & \textbf{5.42}* & \textbf{5.38}* & \textbf{5.67}* \\\hline
\end{tabular}
\end{table}
\begin{table}[t]
\footnotesize
\centering
\caption{The average tensor reconstruction loss.}
\label{tensor_loss}
\begin{tabular}{|l|cc|c|c|cc|}
\hline
           &    \multicolumn{2}{c|}{Twitch-Full}  &  Twitch-Chat    &     Douyu       &   \multicolumn{2}{c|}{Reddit}   \\\cline{2-7}
           &    Don.         &   Resp.            &   Don.          &   Don.          &   Don.          &   Resp.         \\ \hline
 Tucker    &         6.98    &        10.35       &     0.52        &   12.33         &      10.30      &        15.97    \\\hline
\tensorser & \textbf{1.99}*  & \textbf{5.21}*     &       -         &       -         & \textbf{2.67}*  & \textbf{8.11}*  \\
\tensors   & \textbf{2.01}*  & \textbf{4.52}*     & \textbf{0.30}*  & \textbf{4.12}*  & \textbf{2.99}*  & \textbf{8.09}*  \\
\tensorr   & \textbf{1.55}*  & \textbf{6.73}*     &       -         &       -         & \textbf{2.41}*  & \textbf{7.08}*  \\
\tensorn   & \textbf{2.57}*  & \textbf{8.01}*     &       -         &       -         & \textbf{3.97}*  & \textbf{10.38}*  \\
\tensor    & \textbf{1.32}*  & \textbf{3.44}*     & \textbf{0.24}*  & \textbf{3.67}*  & \textbf{2.31}*  & \textbf{6.66}*  \\ \hline
\multicolumn{7}{l}{*: statistically significant (p-value $\leq 0.007<0.05$)}
\end{tabular}
\end{table}

\subsection{Evaluation of Donation Recommendations}\label{exp:donation}
\begin{table*}[t]
\footnotesize
\centering
\caption{Top-K ranking performances of \mwp\ recommendations.}
\label{large_full}
\begin{tabular}{|c|c|c|c|c|c|c|c|c|c|c|c|c|c|c|c|c|c|c|c|c|}
\hline
%----------------------- Twitch ---------------------------
       &   \multicolumn{4}{c|}{Twitch-Full}                                                 & \multicolumn{4}{c|}{Twitch-Chat} & \multicolumn{4}{c|}{Douyu} & \multicolumn{4}{c|}{Reddit} \\ \hline
       &     \multicolumn{2}{c|}{K=2}      &     \multicolumn{2}{c|}{K=4}      &     \multicolumn{2}{c|}{K=2}      &     \multicolumn{2}{c|}{K=4}      &     \multicolumn{2}{c|}{K=2}      &     \multicolumn{2}{c|}{K=4}      &     \multicolumn{2}{c|}{K=2}        &     \multicolumn{2}{c|}{K=4}      \\ \cline{2-17} 
       &        HR       &       MAP       &        HR       &       MAP       &      HR         &       MAP       &        HR       &       MAP       &        HR       &       MAP       &      HR         &       MAP       &        HR       &       MAP       &        HR       &       MAP       \\ \hline
NCF    &     0.306       &      0.297      &      0.529      &      0.400      &      0.209      &      0.165      &     0.389       &      0.237      &      0.277      &      0.223      &      0.351      &      0.288      & 0.303 & 0.289 & 0.503 & 0.354\\ %END NCF
SBPR   &     0.310       &      0.292      &      0.560      &      0.386      &      0.236      &      0.188      &     0.440       &      0.302      &      0.322      &      0.279      &      0.507      &      0.389      & 0.288 & 0.253 & 0.481 & 0.317\\ %END SBPR
AGREE  &     0.312       &      0.299      &      0.566      &      0.407      &      0.241      &      0.189      &     0.439       &      0.289      &      0.345      &      0.281      &      0.561      &      0.422      & 0.306 & 0.271 & 0.514 & 0.366\\ %END AGREE
GBPR   &     0.233       &      0.209      &      0.422      &      0.306      &      0.222      &      0.170      &     0.388       &      0.244      &      0.232      &      0.200      &      0.321      &      0.228      & 0.206 & 0.181 & 0.322 & 0.239 \\\hline %END GBPR
\ranku & \textbf{0.500}* & \textbf{0.443}* & \textbf{0.784}* & \textbf{0.609}* & \textbf{0.352}* & \textbf{0.283}* & \textbf{0.586}* & \textbf{0.402}* & \textbf{0.407}* & \textbf{0.319}* & \textbf{0.728}* & \textbf{0.599}* & \textbf{0.421}* & \textbf{0.388}* & \textbf{0.632}* & \textbf{0.527}*\\ %END ranku
\rankf & \textbf{0.519}* & \textbf{0.456}* & \textbf{0.809}* & \textbf{0.622}* & - & - & - & - & - & - & - & - & \textbf{0.451}* & \textbf{0.389}* & \textbf{0.655}* & \textbf{0.550}* \\ %END rankf
\rankc & \textbf{0.499}* & \textbf{0.412}* & \textbf{0.773}* & \textbf{0.589}* & - & - & - & - & - & - & - & - & \textbf{0.460}* & \textbf{0.391}* & \textbf{0.701}* & \textbf{0.589}*\\ % END rankc
\rankn & \textbf{0.431}* & \textbf{0.378}* & \textbf{0.642}* & \textbf{0.512}* & \textbf{0.314}* & \textbf{0.241}* & \textbf{0.521}* & \textbf{0.371}* & \textbf{0.388}* & \textbf{0.314}* & \textbf{0.654}* & \textbf{0.486}* & \textbf{0.353}* & \textbf{0.323}* & \textbf{0.567}* & \textbf{0.399}*\\ %END rankn
\rank  & \textbf{0.523}* & \textbf{0.484}* & \textbf{0.832}* & \textbf{0.664}*  & \textbf{0.421}* & \textbf{0.372}* & \textbf{0.673}* & \textbf{0.453}* & \textbf{0.499}* & \textbf{0.442}* & \textbf{0.798}* & \textbf{0.608}* & \textbf{0.463}* & \textbf{0.406}* & \textbf{0.752}* & \textbf{0.572}*\\ \hline % END rank
\end{tabular}
\end{table*}
Table \ref{donation_full} shows the prediction performance (RMSE) in Twitch-Full and Reddit.\footnote{\label{no_response} Results of donation recommendations and response tensor reconstruction loss of Twitch-Chat and Douyu are not shown since reciprocal responses are not provided.} \dtor\ outperforms others by at least 78.2\% and 41.9\% in Twitch-Full and Reddit, respectively, indicating that \textit{the proposed features for \dtor} and \textit{\smooth} are effective to estimate reciprocal responses. \dtor\ improves \dtorn\ by at least 62.2\% in Twitch-Full since \textit{the asymmetric viewer and streamer communication behaviors} play a crucial role in live streaming and \smooth\ properly encodes it in embeddings. In contrast, authors in Reddit are rarely overwhelmed with numerous comments (2.3 times less than that in Twitch-Full), and the improvement (25.3\%) is thereby less significant. Similarly, the improvement of \dtor\ over ATS and STNN is smaller in Reddit (41.9\%) than in Twitch-Full (78.2\%). DLR performs the worst, indicating that the donation amount is not the only factor for acquiring high-quality responses. All algorithms perform the best when $L=4$, indicating that 8 and 20 minutes are adequate for modeling socio-temporal behaviors in Twitch-Full and Reddit, respectively. The value is smaller in Twitch-Full since streamers usually respond faster via verbal sentences rather than typing in Reddit.

Table \ref{tensor_loss} compares the reconstruction loss in all datasets.$^\text{\ref{no_response}}$ In Twitch-Full and Reddit, \tensor\ and its variants outperform Tucker Decomposition by at least 171.5\% and 29.2\% with respect to the donation and response tensors, manifesting that the task of co-factorizing donation and response tensors helps each other by \textit{sharing latent matrices}. Note that \tensorser\ reconstructs both tensors egregiously in Twitch-Full (dropping by 50.7\% and 51.4\% compared with \tensor) since streamer relations are critical in multi-streaming. In Twitch-Full, the performance of \tensorr\ in the response tensor is worse than \tensor\ (95.6\%), demonstrating that \smooth\ is effective in modeling the asymmetric behaviors. In contrast, \tensor\ merely improves \tensorr\ for donation tensor reconstruction in Reddit (4.3\%) since the asymmetric behaviors are not obvious (consistent with the observations in prediction performances). Moreover, \tensors\ shows a weaker impact for donation tensor reconstruction in Twitch-Chat (20\%) and Douyu (10.9\%) than in Twitch-Full (52.2\%), because Douyu supports \textit{one-key gifting}, which allows viewers to send specified gifts quickly by pressing a predefined button, and viewers tend to separate their budgets to send multiple small gifts to attract the streamers (94.6\% of consecutive gifts sent within 60 seconds in a channel). The ripple effect is thereby diluted by dense and consecutive donations. Similarly, viewers keep sending chats since it is free, and those frequent chats also cause a non-obvious ripple effect in Twitch-Chat.

\subsection{Evaluation of \mwp\ Recommendations}\label{exp:mwp}
\begin{figure}[tp]
	\centering
	\subfigure[][Twitch-Full.] {\
		\centering \includegraphics[width = 0.45 \columnwidth] {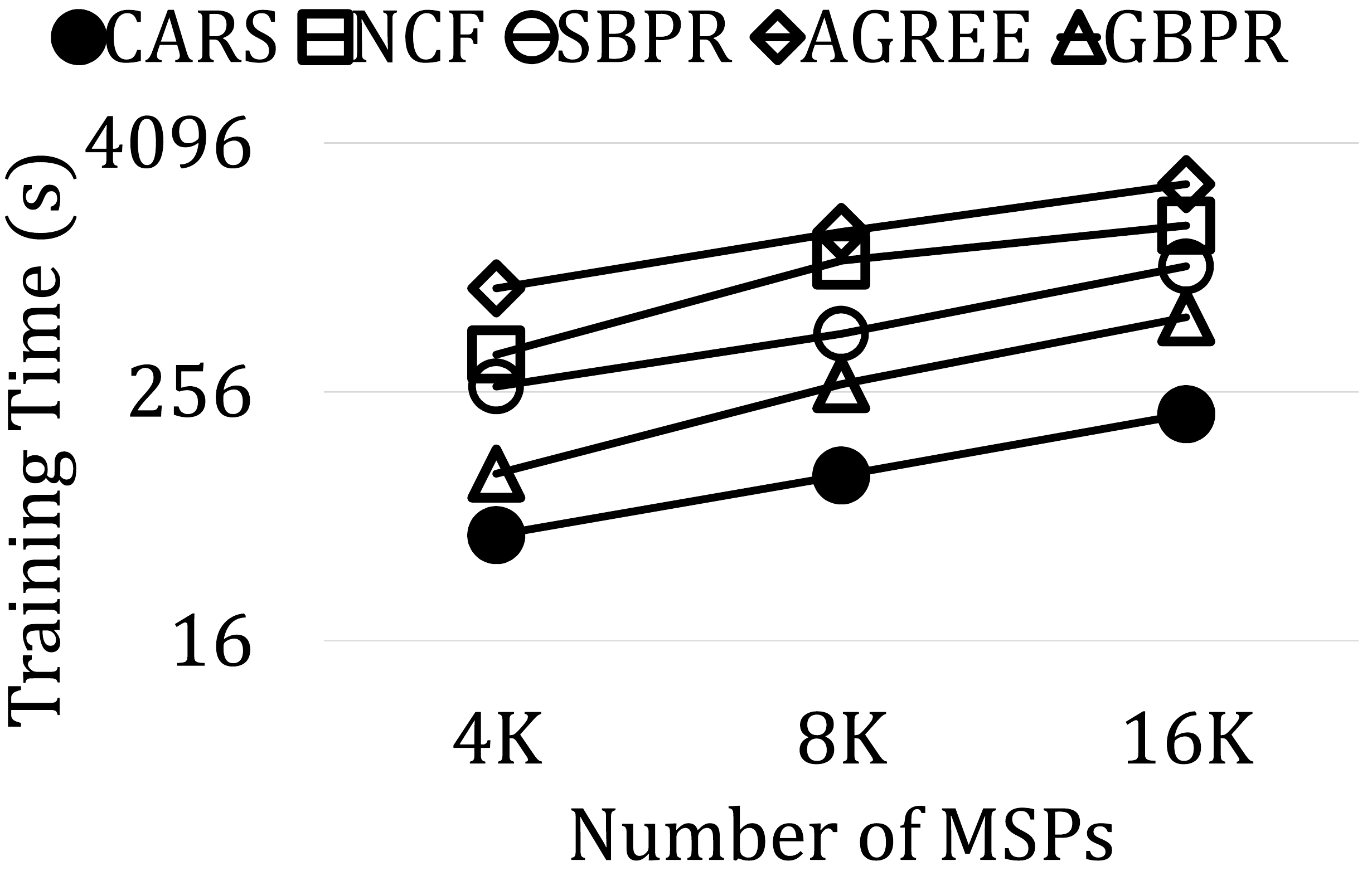}
		\label{fig:msp_twitch_time} } 
	\subfigure[][Reddit.] {\ \centering  \includegraphics[width = 0.45 \columnwidth] {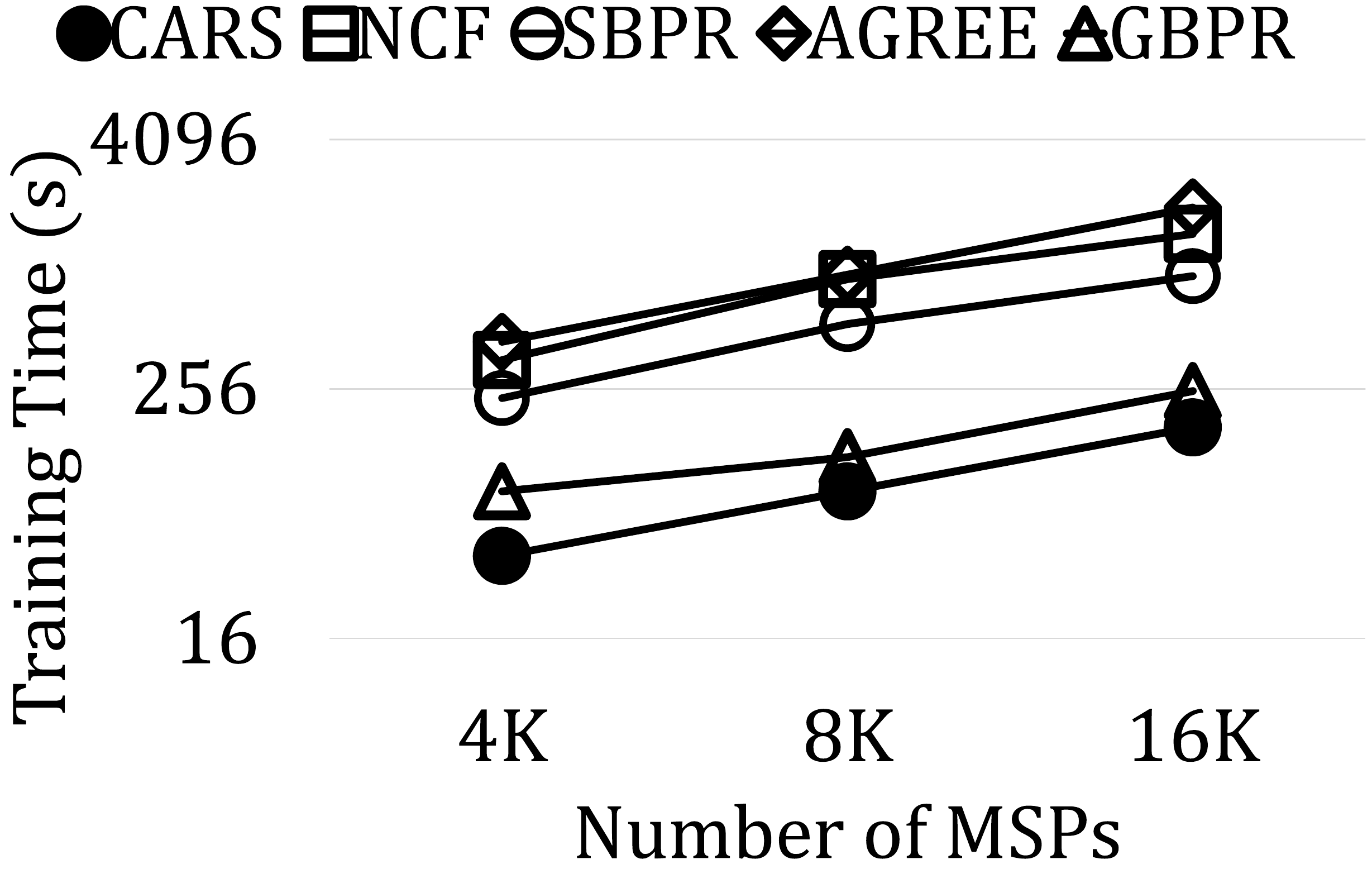}
		\label{fig:time_reddit_time} }
	\caption{Training time for ranking on different datasets.}
	\label{fig:rank_time}
\end{figure}

Table \ref{large_full} presents the results of HR@K and MAP@K with different $k$, where \rankf\ and \rankc\ are left blank on Twitch-Chat and Douyu since they do not provide social networks of viewers and streamers. \rank\ outperforms other approaches by at least 38.8\% and 40.4\% regarding HR and MAP, respectively. \framework\ successfully integrates \tensor\ and \rank\ by ranking \mwp s based on the embeddings extracted from factorization results. In Twitch-Full, \rank\ outperforms state-of-the-art recommenders (HR: 46.9\%, MAP: 61.8\%) since it parameterizes \textit{channel influence} for each viewer-channel pair to strike a good balance between personal interests, social interactions, and streamer relations. Moreover, \rankn\ also outperforms other baselines (at least 11.7\%) in live streaming datasets, indicating the above improvement comes not only from the pre-trained embeddings but also channel influence. The performance of \rankc\ plummets the most among all variants in Twitch-Full but the improvement is less in Reddit. It is because viewers enjoy watching streamers collaborating or competing in multi-streaming, while the authors of posts (streamers) in Reddit interact less with each other. Compared with \ranku\ and \rankf, the performance of \rankn\ drops in both Twitch-Full and Reddit by at least 28.0\% regarding MAP@2, demonstrating that the embeddings extracted by \tensor\ are effective. \rank\ performs better in Twitch-Full than in Twitch-Chat and Douyu since \tensor\ incorporates meaningful donation-response relations by co-factorization. Note that NCF performs better in Reddit than in the others because individual user behaviors are more dominant in Reddit, evident by a smaller weight of the social aspect $\tau^V_v$ of channel influence in Reddit (0.17) and in Twitch-Full (0.29). %AGREE outperforms NCF in Douyu (HR@4: 59.8\%) than in Reddit (HR@4: 3.7\%) indicating that both personal interests and group interactions are important. 

\noindent \textbf{Running time.} Figure \ref{fig:rank_time} shows the training time with different $k$. The training time of \rank\ is faster than others since it does not train the viewer and channel embeddings. NCF and AGREE are significantly slower due to their deep neural network structure. Note that \rank\ takes only $10^{-5}$ seconds to evaluate 1 \mwp\ in the inference stage, and is thereby suitable for online scenarios.

\subsection{Analysis of Social Influence}\label{exp:social_influence}
\begin{figure}[tp]
	\centering
	\subfigure[][$\mathbf{W}_V$ before training.] {\	\centering \includegraphics[width = 0.40 \columnwidth] {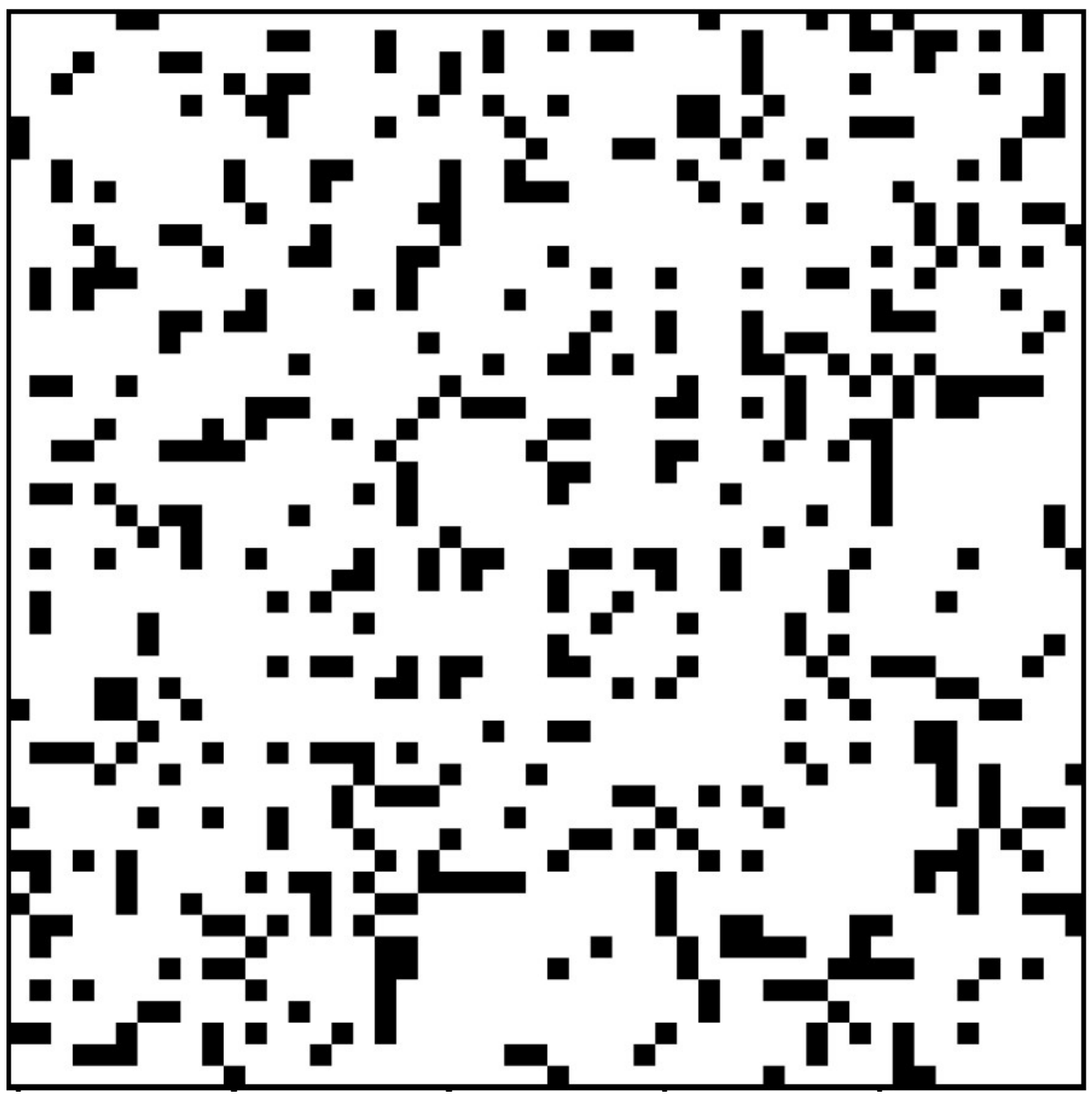}
		\label{fig:original_adj} } 
	\subfigure[][$\mathbf{W}_V$ after training.] {\ \centering  \includegraphics[width = 0.45 \columnwidth] {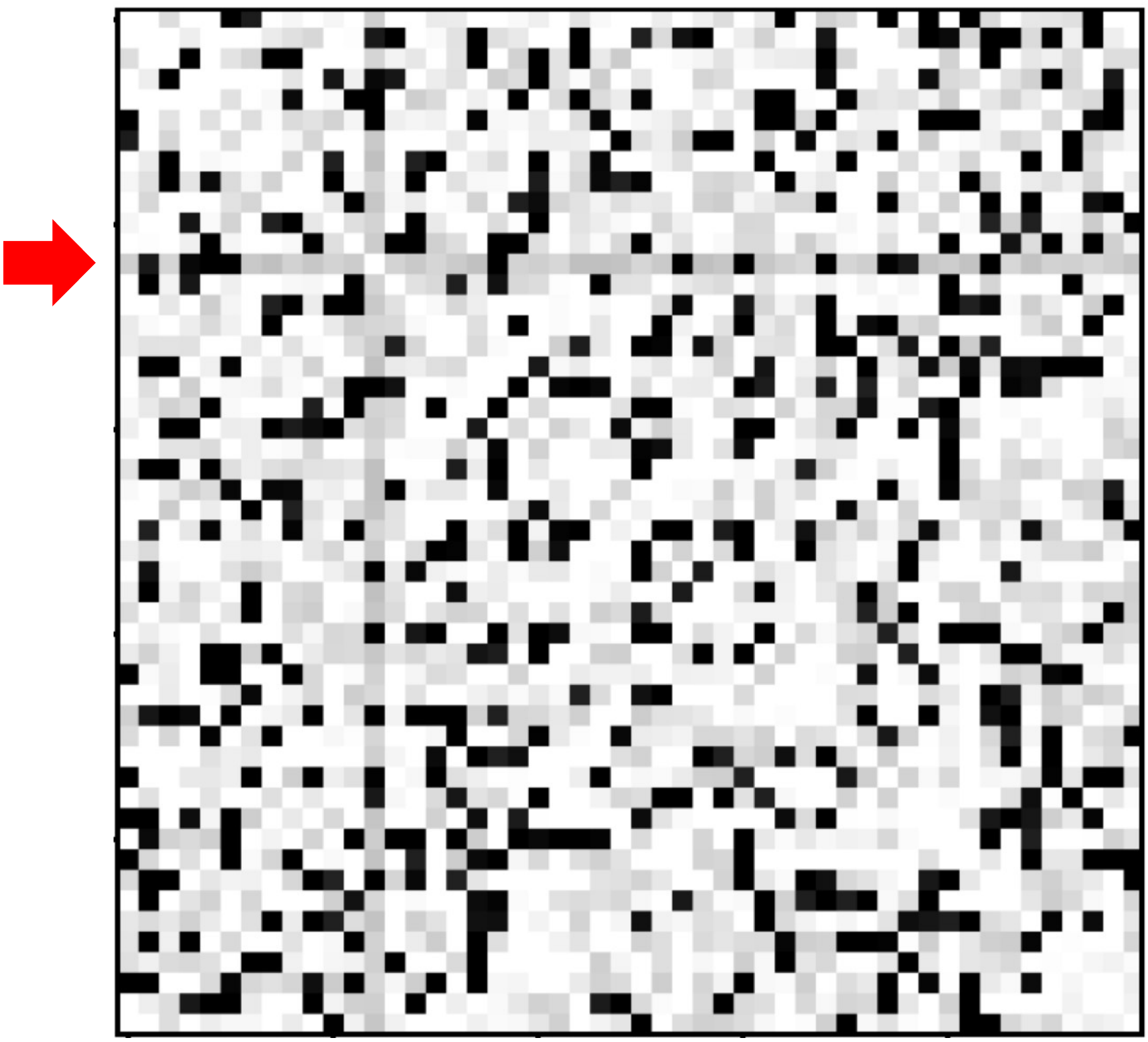}
		\label{fig:weighted_adj} }
	\caption{Social influence matrix before and after training.}
	\label{fig:social_influence}
\end{figure}

Figure \ref{fig:social_influence} compares the partial social influence matrix $\hat{\mathbf{W}_V}$ of 50 viewers (for clarity) before and after training in Twitch-Full. The gray-level of an entry $\hat{\mathbf{W}_V}(i,j)$ is closer to black if the social influence between viewers $i$ and $j$ is stronger. In Figure \ref{fig:original_adj}, $\hat{\mathbf{W}_V}(i,j)$ is initialized to 1 if the viewers are friends ($(i,j)\in E$) and $\epsilon=0.02$ otherwise. After training, many gray pixels appear, manifesting that even non-friend viewers may trigger socio-temporal ripple effects on donations. $\hat{\mathbf{W}_V}$ thus improves \dtor\ from \dtor -without-\ripple\ (RMSE: 40.1\%), and \rank\ from \rankf\ (HR@4: 2.8\%, MAP@4: 6.7\%). Note that one special viewer, who is a famous streamer in the LOL community, influences every other viewer (pointed by the red arrow) but is not a friend of all viewers. His donation on other channels easily causes sensations since ordinary viewers are excited to see social interactions between their favorite streamers, and more donations are thereby triggered. In summary, \tensor\ is able to quantify social influence based on socio-temporal donation behaviors to improve donation and \mwp\ recommendations.

\begin{table}[t]
\footnotesize
\centering
\caption{Performances (RMSE) of different topics of channels in Twitch-Full on donation recommendations.}
\label{donation_games}
\begin{tabular}{|c|c|c|c|c|}
\hline
      &      LOL       &    Fortnite    &      CS: GO     & Just Chatting  \\\hline 
DLR    &         5.91   &         6.88   &         5.66   &         4.33   \\
ATS    &         3.66   &         3.87   &         3.52   &         3.61   \\
STNN   &         3.53   &         3.67   &         3.59   &         3.57   \\\hline
\dtorn & \textbf{2.77}* & \textbf{2.92}* & \textbf{2.53}* & \textbf{2.45}* \\
\dtor  & \textbf{1.68}* & \textbf{1.73}* & \textbf{1.44}* & \textbf{2.01}* \\\hline
\end{tabular}
\end{table}

\subsection{Comparison of Different Communities}\label{exp:diff_channel}
Table \ref{donation_games} compares the performances of all methods in 4 different communities of channels in Twitch-Full. \dtor\ outperforms the baselines by at least 52.7\% in every category. However, the improvement of \dtor\ over \dtorn\ is smaller in Just Chatting (21.9\%) than others (at least 64.8\%). To discover the insights, we respectively select two famous streamers (>3K viewers) from gaming (LOL) and Just Chatting. We observe that donations burst when some game events happen. For example, when the streamer of LOL makes stunning plays, viewers donate and send messages like ``Coooooool!!!!'' and ``Dude, that's insane'' for the nice plays. To face a huge amount of immediate donations, the streamer says ``Yes! I've told you guys to trust me! Thanks for your donations by the way!'' to all the donors with short reciprocal responses and continue the game playing. In contrast, bursts of donations in Just Chatting are rarely observed. Viewer donations and streamer responses are usually orderly, i.e., the viewers are inclined to donate to the streamer when she is free to respond. Consequently, the effectiveness of \smooth\ in gaming channels (e.g., LOL) is more evident than in Just Chatting.

\section{Conclusion}
To the best of our knowledge, we make the first attempt to exploit unique phenomena of multi-stream and donations in live streaming. In this paper, we formulate \prob\ and propose a two-phase framework \framework. The novelty of MARS lies under the design of its components, including i) \tensor\ extracts discriminative features by jointly learning viewer and streamer behaviors, ii) \dtor\ recommends streamers to donate to for viewers, and iii) \rank\ learns channel influence to rank \mwp s. Experimental results manifest that \dtor\ significantly outperforms other feedback prediction models by at least 41.9\% in terms of root-mean-square error, and \rank\ significantly outperforms personalized and group recommendations by at least 38.8\%  and 40.4\% in large datasets in terms of hit ratio and mean average precision.

\section*{Acknowledgment}
This work is supported in part by NSF under grants IIS-1717084, by MOST in Taiwan through grant 108-2221-E-009-088, 109-2221-E-001-015, 107-2221-E-001-011-MY3, 108-2221-E-001-002, 109-2221-E-001-017-MY2, 109-2218-E-009-015, and 109-2221-E-009-118-MY3. We thank National Center for High-performance Computing (NCHC) of National Applied Research Laboratories (NARLabs) in Taiwan for providing computational and storage resources.

\bibliographystyle{ACM-Reference-Format}
\bibliography{new_ref} 

\end{document}